\documentclass[journal]{IEEEtran}
\usepackage{commath,graphicx,afterpage,subfigure,amsmath,amsfonts,amssymb,algorithm,algorithmic}
\usepackage{graphicx,times,amssymb,cite,subfigure,stfloats,booktabs,multirow,bm, pifont}
\usepackage{cite,multirow,bm,multicol,booktabs,threeparttable,extpfeil}
\usepackage{graphicx}

\begin{document}

\title{MVCNet: Multi-View Contrastive Network for \\ Motor Imagery Classification}

\author{Ziwei~Wang, Siyang~Li, Xiaoqing~Chen, and Dongrui~Wu$^{\ast}$, \IEEEmembership{Fellow,~IEEE}
\thanks{This research was supported by Shenzhen Science and Technology Innovation Program JCYJ20220818103602004, and Zhongguancun Academy 20240301.}
\thanks{Z.~Wang, S.~Li, X.~Chen, and D.~Wu are with the Ministry of Education Key Laboratory of Image Processing and Intelligent Control, School of Artificial Intelligence and Automation, Huazhong University of Science and Technology, Wuhan 430074, China. They are also with the Shenzhen Huazhong University of Science and Technology Research Institute, Shenzhen, 518000 China.}
\thanks{*Corresponding Author: Dongrui Wu (drwu09@gmail.com).}}

\markboth{}
{Shell \MakeLowercase{Wang~\emph{et al.}}: }
\maketitle

\begin{abstract}
Electroencephalography (EEG)-based brain-computer interfaces (BCIs) enable neural interaction by decoding brain activity for external communication. Motor imagery (MI) decoding has received significant attention due to its intuitive mechanism. However, most existing models rely on single-stream architectures and overlook the multi-view nature of EEG signals, leading to limited performance and generalization. We propose a multi-view contrastive network (MVCNet), a dual-branch architecture that parallelly integrates CNN and Transformer blocks to capture both local spatial-temporal features and global temporal dependencies. To enhance the informativeness of training data, MVCNet incorporates a unified augmentation pipeline across time, frequency, and spatial domains. Two contrastive modules are further introduced: a cross-view contrastive module that enforces consistency of original and augmented views, and a cross-model contrastive module that aligns features extracted from both branches. Final representations are fused and jointly optimized by contrastive and classification losses. Experiments on five public MI datasets across three scenarios demonstrate that MVCNet consistently outperforms nine state-of-the-art MI decoding networks, highlighting its effectiveness and generalization ability. MVCNet provides a robust solution for MI decoding by integrating multi-view information and dual-branch modeling, contributing to the development of more reliable BCI systems.
\end{abstract}


\section{Introduction}
A brain-computer interface (BCI) serves as a direct communication pathway between a user's brain and an external device \cite{Rosenfeld2017}. BCIs have a crucial role in mapping, assisting, augmenting, and potentially restoring human cognitive and/or sensory-motor functions \cite{Krucoff2016}. Furthermore, BCIs contribute significantly to cognitive behavior assessment, pain management, emotional regulation, neurogaming, etc \cite{van2012brain}. BCIs can be categorized into non-invasive, partially invasive, and invasive ones, based on the proximity of electrodes to the brain cortex \cite{wu2020transfer}. Among them, non-invasive electroencephalography (EEG)-based BCIs stand out due to their convenience and cost-effectiveness.

Motor imagery (MI) \cite{Pfurtscheller2001} paradigm involves users imagining the movement of specific body parts (e.g., left hand, right hand, both feet, or tongue), modulating different regions of the brain’s motor cortex \cite{Wu2022NN}. Despite substantial progress, decoding MI from EEG signals remains challenging. The intrinsic non-stationarity, low signal-to-noise ratio, and substantial inter-subject variability of EEG data impose significant barriers to robust MI decoding.

Various networks have been proposed for MI decoding. Representative approaches based on the convolutional neural network (CNN), such as EEGNet \cite{Lawhern2018EEGNet} and SCNN \cite{deepshallow2017}, are proved effective. However, CNNs are inherently limited in modeling global temporal dependencies, which are essential for MI tasks involving sequential mental processes. To address this, recent works have explored hybrid CNN-Transformer architectures. For example, EEG Conformer \cite{song2022eeg} integrates a CNN block and a self-attention block in a sequential manner, where the CNN block extracts spatial-temporal representations, and the self-attention module further refines long-term temporal features. Nevertheless, such serial designs restrict the interaction between local and global feature representations.

\begin{figure}[htpb] \centering
\includegraphics[width=\linewidth,clip]{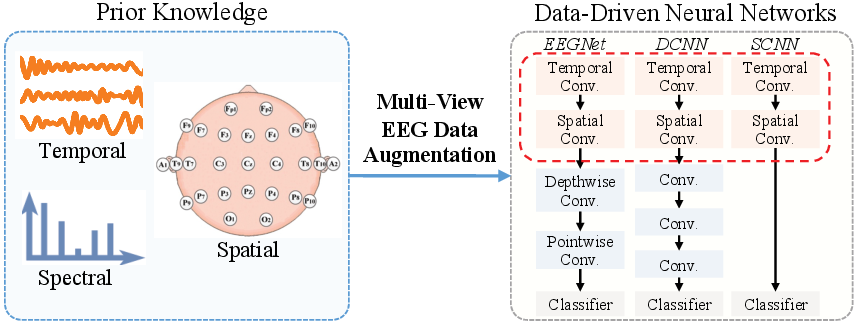}
\caption{Illustration of integrating prior knowledge into data-driven neural networks with data augmentation, e.g., temporal, spatial, and/or spectral information. Multiple views of the EEG data ensure that feature learning surpasses the limitation of designated networks. As an example, EEGNet, DCNN, and SCNN architectures rely on the intuition of CSP or filter-bank CSP for spatial variance maximization across classes (red dotted box), while not all important characteristics from time, spatial, and frequency domains of EEG signals are investigated.} \label{fig:intro}
\end{figure}

A further limitation of current MI decoding networks lies in their reliance on single-view representations. Most work focuses only on raw EEG signals, while others, like FBCNet \cite{mane2021fbcnet} and IFNet \cite{wang2023ifnet}, construct multiple filter bands to obtain various spectral views. Yet, few approaches systematically explore and fuse complementary views across time, frequency, and spatial domains. The lack of diversity hinders the model's generalization ability, especially in cross-subject \cite{Li2024T-TIME} and cross-headset \cite{wang2025cst} scenarios. Thus, developing principled decoding models that fuse multi-view information and encode inductive priors remains a pressing need. Multi-view prior knowledge can be integrated into data-driven MI decoding networks through data augmentations, as illustrated in Figure~\ref{fig:intro}.

To address above limitations, we propose MVCNet, a multi-view contrastive decoding network tailored for MI decoding. MVCNet features a dual-branch parallel architecture comprising CNN and Transformer branches, enabling the concurrent extraction of both localized spatial-temporal features and long-range temporal dependencies. Furthermore, we introduce cross-view and cross-model contrastive regularization to enhance alignment and discrimination across augmented views and network branches. To better leverage the diverse nature of EEG data, we design a multi-view data augmentation strategy that jointly considers time, frequency, and channel transformation, effectively injecting domain priors and boosting generalization.

The main contributions of this work are:
\begin{itemize}
\item We propose MVCNet, a dual-branch parallel network integrating CNN and Transformer models for MI decoding. This architecture effectively captures both local spatial-temporal features and global temporal dependencies within EEG data.
\item We introduce the cross-view contrasting (CVC) and cross-model contrasting (CMC) modules to enhance feature alignment and discriminability across diverse views and network branches.
\item We design a multi-view EEG data augmentation strategy that operates across time, frequency, and spatial domains, incorporating informative priors to improve model generalization.
\item Extensive experiments on five public MI datasets across three scenarios demonstrate that MVCNet outperforms nine state-of-the-art MI decoding networks and seven data augmentation strategies, validating its effectiveness and robustness.
\end{itemize}

The remainder of this paper is organized as follows: Section~\ref{sect:rw} reviews related works. Section~\ref{sect:me} details the proposed MVCNet. Section~\ref{sect:er} discusses the experimental results and provides analyses. Section~\ref{sect:conclusions} draws conclusions.

\section{Related Work}\label{sect:rw}
\subsection{MI Decoding Networks}
EEG-based MI decoding has garnered substantial attention over the past decades. Researchers have concentrated on developing diverse network architectures to enhance MI decoding performance. Increasingly, end-to-end deep neural networks, spanning CNN-based, Transformer-based, and the more recent Mamba-based architectures, have been widely adopted and evaluated:
\begin{itemize}
\item CNN-based models: CNNs remain the most prevalent model for EEG decoding due to their efficacy in capturing local spatial-temporal patterns. EEGNet \cite{Lawhern2018EEGNet} is among the most widely used models, featuring two convolutional blocks and a classification head. DCNN \cite{deepshallow2017} employs four convolutional blocks with more parameters. SCNN \cite{deepshallow2017}, a simplified variant of DCNN, is inspired by the filter bank common spatial pattern approach. FBCNet \cite{mane2021fbcnet} extracts spectral-spatial representations by integrating spatial convolution and temporal variance layers. Based on FBCNet, FBMSNet \cite{liu2022fbmsnet} incorporates extended convolutions for multi-scale feature extraction and introduces center loss to align features with class centers. ADFCNN \cite{tao2023adfcnn} proposes a dual-branch CNN structure coupled with a self-attention module to enhance inter-branch feature fusion. IFNet employs two convolutional layers to extract spectral-spatial representations, similar to FBCNet. Ma \textit{et al.} \cite{ma2023temporal} propose a comparable model integrating FBCNet with a temporal attention layer.

\item Transformer-based models: Transformers have been introduced into EEG decoding owing to their strong capability in modeling global temporal dependencies. EEG Conformer \cite{song2022eeg} adopts a sequential architecture combining SCNN with a Transformer block to capture high-level spatial-temporal patterns. Zhao \textit{et al.} \cite{zhao2024ctnet} utilize a convolutional module, similar to EEGNet, to extract local and spatial features, and a Transformer encoder to discern global dependencies.

\item Mamba-based models: Mamba \cite{gu2023mamba}, a recently introduced alternative to Transformers, addresses the challenges of long-sequence modeling and resource inefficiencies. EEGMamba \cite{gui2024eegmamba} utilizes a bidirectional Mamba module to  capture dependencies across EEG tokens. MI-Mamba \cite{guo2025mi} integrates CNN and Mamba blocks sequentially, akin to the architecture of EEG Conformer. SlimSeiz \cite{lu2024slimseiz} combines multiple 1D CNN layers and a Mamba block to simultaneously extract temporal features at various resolutions and long-range temporal dependencies for seizure prediction.
\end{itemize}

Building upon prior work that models EEG trials through spectral, spatial, and temporal perspectives to extract salient features, we propose a parallel dual-branch architecture that integrates CNN and Transformer models. This design leverages the complementary strengths of CNNs in capturing local spatial-temporal patterns and Transformers in modeling global temporal dependencies, enabling more comprehensive and effective feature representation.

Recent studies have explored the integration of domain-specific priors to improve MI decoding performance. For example, temporal, spatial, or spectral characteristics of EEG signals have been explicitly encoded through data augmentation \cite{Wang2024,Wang2025CSDA}, time-frequency transformations \cite{dong2025noise}, frequency band modeling \cite{wang2023ifnet,liu2022fbmsnet}, initialization with spatial filters \cite{jiang2024csp}, and multi-species/modality information fusion \cite{wang2025cst}. These approaches demonstrate the effectiveness of knowledge-data fusion decoding. However, most existing approaches incorporate such priors independently within specific modules or modalities. In contrast, our work unifies diverse temporal, spectral, and spatial priors into a joint learning framework via multi-view contrastive learning, thereby enabling more comprehensive representation learning and improved generalization.

\subsection{EEG Data Augmentation}
EEG data augmentation serves as an effective solution to improve model generalization, particularly in scenarios with limited data or a large distribution discrepancy. Existing EEG data augmentations typically operate in the time, frequency, or spatial domains, aiming to generate plausible variations of EEG trials while preserving their semantic consistency.

In the time domain, Wang \textit{et al.} \cite{Wang2018} introduced random Gaussian white noise to augment temporal variability. Mohsenvand \textit{et al.} \cite{Mohsenvand2020} proposed time-masking strategies by zeroing out random segments of EEG trials, and Rommel \textit{et al.} \cite{Rommel2021} applied trial-level temporal flips or reversed the time axis across all channels to perturb signal directionality. In the frequency domain, Schwabedal \textit{et al.} \cite{Schwabedal2018surr} randomized the phase components of EEG signals in the Fourier space to produce surrogate data, while Mohsenvand \textit{et al.} \cite{Mohsenvand2020} and Cheng \textit{et al.} \cite{cheng2020subject} introduced selective filtering of narrow-band frequency components to simulate spectral shifts. Rommel \textit{et al.} \cite{Rommel2021} further perturbed the power spectral density to emulate variations in signal power distribution. Spatial domain augmentations aim to diversify channel-level patterns. Wang \textit{et al.} \cite{Wang2024} exchanged the symmetrical left and right hemisphere channels, as well as labels in the left/right hand MI task. Saeed \textit{et al.} \cite{Saeed2021} explored channel dropout and permutation to simulate electrode variability. Krell \textit{et al.} \cite{Krell2017} introduced channel interpolation over randomly rotated topographies, and Pei \textit{et al.} \cite{pei2021hs} performed hemispheric recombination by swapping left and right brain regions from different samples, thus enhancing spatial diversity across brain regions.

However, existing approaches usually consider single augmentation strategy at a time, without integrating multi-view knowledge. As a result, the augmented representations often fail to fully capture the complex, multi-faceted nature of EEG data across time, frequency, and space domains.

To address this limitation, we design a multi-view contrastive framework, where diverse augmentations are performed to construct semantically consistent but statistically diverse views of EEG data. These views are then aligned via a dedicated CVC module. By leveraging diverse signal characteristics from multiple views, MVCNet promotes the learning of multi-view expressive features.

\subsection{Contrastive Learning}
Contrastive learning has emerged as a powerful way to learn discriminative representations by encouraging similar instances (positive pairs) to be mapped close together while pushing dissimilar ones (negative pairs) farther apart in the feature space.

Most work focuses on unsupervised representation learning. SimCLR \cite{simclr2020} constructs two augmented views of each input and enforces consistency between them using the normalized temperature-scaled cross-entropy (NT-Xent) loss. In SimCLR, augmented samples derived from the same input are treated as positives, while all others within the batch are considered negatives. MoCo \cite{he2020moco} improves scalability by introducing a dynamic memory bank to maintain a large queue of negative examples, and employs a momentum encoder to stabilize training. Some works explore approaches that do not rely explicitly on negative samples. BYOL \cite{grill2020BYOL} designs a dual-network setup, where one network predicts the output of another, both learning collaboratively via a bootstrapping mechanism. SimSiam \cite{chen2021simsiam} further simplifies the setup by eliminating both negative pairs and momentum encoders, instead relying on a stop-gradient operation within a siamese architecture to prevent representational collapse. Compared to the unsupervised setting, supervised contrastive learning has received relatively limited attention. Nevertheless, it offers unique advantages by leveraging label information to construct semantically meaningful positive and negative pairs. For instance, SupCon \cite{Khosla2020} extends the SimCLR framework by considering all samples sharing the same class label as positives, and the remaining samples within the batch as negatives.

Contrastive learning has been explored to improve representation quality in BCIs. Cheng \textit{et al.} \cite{cheng2020subject} incorporated a subject-specific contrastive loss and adversarial training to promote subject-invariant feature learning. In seizure detection, Huang \textit{et al.} \cite{huang2023epilepsynet} employed contrastive objectives to mitigate the dependency on large-scale labeled datasets. For sleep stage classification, Jiang \textit{et al.} \cite{jiang2021self} introduced a transformation-discrimination pretext task, while Lee \textit{et al.} \cite{lee2022self} enhanced the quality of contrastive pairs via attention mechanisms. Mohsenvand \textit{et al.} \cite{Mohsenvand2020} extended SimCLR to the EEG domain with a channel-wise feature extractor tailored for spatio-temporal signals. Zhang \textit{et al.} \cite{zhang2022expert} integrated expert knowledge to refine local and contextual representations, while Weng \textit{et al.} \cite{weng2023knowledge} leveraged neurological priors to guide contrastive representation learning. Most existing contrastive learning approaches in EEG decoding remain unsupervised and thus heavily dependent on large volumes of data, as well as extensive hyperparameter tuning. These limitations can hinder their practical applicability in supervised BCI tasks such as motor imagery decoding, where class labels are available and should be fully utilized.

To tackle the above issues, we introduce two contrastive modules of CVC and CMC to align representations from different views and network branches, respectively. By jointly optimizing two contrastive losses and a conventional cross-entropy loss, our approach enforces feature consistency while maintaining discriminative capacity.

\section{Methodology} \label{sect:me}
\subsection{Overview}
\begin{figure*}[htpb] \centering
\includegraphics[width=\linewidth,clip]{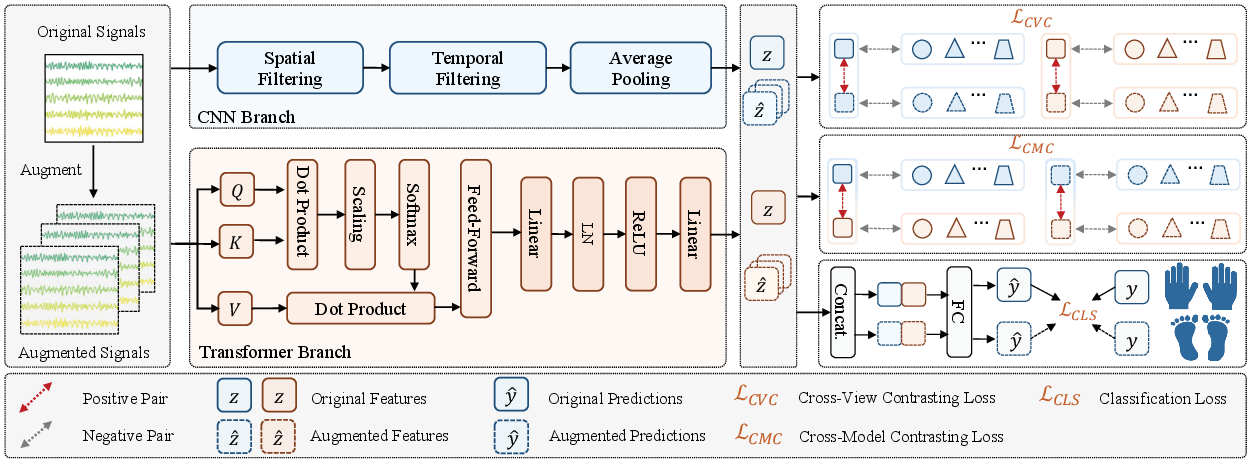}
\caption{The proposed architecture of multi-view contrastive network (MVCNet) for MI decoding.} \label{fig:Framework}
\end{figure*}

This section details the proposed MVCNet. Unlike conventional single branch MI decoding models, such as CNN-based architectures (e.g., EEGNet, DCNN) or serial CNN-Transformer models (e.g., EEG Conformer) that might not capture the full spectrum of EEG characteristics, MVCNet is designed to fully exploit the complementary strengths of convolutional and attention mechanisms via a parallel structure. As illustrated in Figure~\ref{fig:Framework}, the input EEG trials are first augmented to generate multiple views, and then the original and augmented trials are fed into two parallel branches:
\begin{itemize}
\item \textit{CNN branch:} The CNN branch performs spatial filtering, temporal filtering, and average pooling to capture localized spatial-temporal features.
\item \textit{Transformer branch:} The Transformer branch adopts a multi-head self-attention mechanism and feed-forward layers to model global temporal dependencies.
\end{itemize}

To effectively align the cross-view and cross-model representations, two contrastive learning modules and dual-branch feature concatenation are introduced:
\begin{itemize}
\item \textit{Cross-view contrasting:} CVC module enforces consistency between features extracted from original and augmented views of the same sample, while ensuring separation across different samples. The objective is to align representations across augmented views.
\item \textit{Cross-model contrasting:} CMC module aligns features of the same sample obtained from different branches, and contrasts them with features of other samples to maximize inter-sample discrimination. The goal is to promote consistent feature learning across the two branches.
\item \textit{Dual-branch feature fusion}. The features from both branches are concatenated and passed through a fully connected classification head to obtain the final predictions.
\end{itemize}

By integrating multi-view data augmentation, parallel dual-branch feature extraction, and cross-view/model contrastive learning strategies, MVCNet is capable of learning more comprehensive and representative EEG features. The entire framework is trained in an end-to-end manner through the joint optimization of the cross entropy loss and two contrastive losses.

\subsection{Network Architecture}
The architecture of MVCNet is summarized in Table~\ref{tab:network}, which outlines the composition of each module, including layers, kernels, the number of trainable parameters, and output shapes.

The CNN branch is designed to capture localized spatial-temporal dependencies features. It comprises a spatial convolutional layer, a temporal convolutional layer, and an average pooling operation. The spatial convolution layer adopts a depthwise separable 1D convolution equipped with $F$ spatial filters, followed with batch normalization. The temporal convolution layer applies $F$ temporal filters of size $63$, consistent with IFNet \cite{wang2023ifnet}, and is accompanied by batch normalization and a GELU activation. Subsequently, average pooling is employed to reduce the temporal resolution and compress feature dimensionality.

The Transformer branch is configured to model temporal contextual dependencies. It consists of a lightweight self-attention encoder comprising a Transformer block with 2 heads and 2 layers, followed by a two-layer MLP equipped with batch normalization and ReLU activation, following the design in \cite{zhang2022tfc}. A linear projection layer is then applied to normalize the output features into a common latent space, ensuring dimensional consistency with the CNN branch. This branch setup enables the network to learn long-range interactions and higher-order temporal patterns beyond local receptive fields, while facilitating effective feature alignment in the contrastive learning stage.

Finally, the features from both branches are concatenated and passed through a fully connected classification head with a linear activation to obtain the final predictions.

\begin{table*}[htbp]
\centering\setlength{\tabcolsep}{1.5mm}
\small
\renewcommand{\arraystretch}{1.1}
\caption{MVCNet Architecture}
\label{tab:network}
\begin{tabular}{llcccccl}
\toprule
\textbf{Module} & \textbf{Layer} & \textbf{\# Kernels} & \textbf{Kernel Size} & \textbf{\# Parameters} & \textbf{Output Shape} & \textbf{Options} \\
\midrule
\multirow{7}{*}{CNN Branch}
    & Separable Conv1D & $F$ & $(1,)$ & $C \cdot F$ & $(B, F, T)$ & Spatial filtering (depthwise) \\
    & BatchNorm1D &  &  & $2F$ & $(B, F, T)$ &  \\
    & Temporal Conv1D & $F$ & $(64,)$ & $64 \cdot F$ & $(B, F, T)$ & Temporal filtering, padding = same \\
    & BatchNorm1D &  &  & $2F$ & $(B, F, T)$ &  \\
    & GELU &  &  & 0 & $(B, F, T)$ & Nonlinearity \\
    & Average Pooling &  & $(1, W)$ & 0 & $(B, F, T/W)$ & Temporal downsampling \\
    & Dropout &  &  & 0 & $(B, F, T/W)$ & $p=0.5$ \\
\midrule
\multirow{7}{*}{Transformer Branch}
    & Transformer Encoder &  &  & $\sim$ & $(B, T, D)$ & 2 layers, 2 heads, batch-first \\
    & Flatten &  &  & 0 & $(B, T \cdot D)$ & Reshape for projection \\
    & Linear Layer &  &  & $T \cdot D \rightarrow d_1$ & $(B, d_1)$ & Dimensionality reduction \\
    & LayerNorm &  &  & $d_1$ & $(B, d_1)$ & Feature normalization \\
    & ReLU &  &  & 0 & $(B, d_1)$ & Nonlinearity \\
    & Linear Layer &  &  & $d_1 \rightarrow d_p$ & $(B, d_p)$ & Final projection \\
    & Tanh &  &  & 0 & $(B, d_p)$ & Output normalization \\
\midrule
\multirow{2}{*}{Classification}
    & Concatenation &  &  & 0 & $(B, 2d_p)$ & Feature fusion from two branches \\
	& Classifier & $N$ & & $2d_p \cdot N$ & $(B, N_c)$ & Linear fully connected layer \\
\bottomrule
\end{tabular}

\vspace{0.5em}
\footnotesize{
$B$: batch size,
$C$: number of EEG channels,
$T$: number of time points,
$F$: number of CNN filters,
$D$: Transformer embedding dimension,
$W$: pooling window size,
$d_1$: intermediate projector dimension,
$d_p$: final projected feature dimension,
$N_c$: number of classes.
}
\end{table*}

\subsection{EEG Data Augmentation}\label{sub:da_intro}
Seven data augmentation strategies from three views for MI EEG trials were compared and utilized, including three time domain, two frequency domain, and two spatial domain approaches.
\begin{enumerate}
\item Data flipping (Flip) \cite{Zhang2022MSDT}, which flips the EEG trial in the time domain, resulting in opposite voltage values.
\item Noise adding (Noise) \cite{Zhang2022MSDT}, which adds uniform noise to each EEG trial.
\item Data multiplication (Scale) \cite{Zhang2022MSDT}, which multiplies the original EEG trial by a coefficient around 1.
\item Frequency shift (FShift) \cite{Zhang2022MSDT}, which uses Hilbert transform to shift the frequency of EEG trials.
\item Frequency Surrogate (FSurr) \cite{Schwabedal2018surr}, which replaces the Fourier phases of trials by new random numbers from the interval, and applies the inverse Fourier transform.
\item Channel Reflection (CR) \cite{Wang2024}, which exchanges the symmetrical left and right hemisphere channels, as well as the labels.
\item Half Sample (HS) \cite{pei2021hs}, which randomly selects the left brain part and the right brain part of different EEG trials, then recombines the two parts to form a new sample.
\end{enumerate}

\begin{figure*}[htpb]\centering
\includegraphics[width=.9\linewidth,clip]{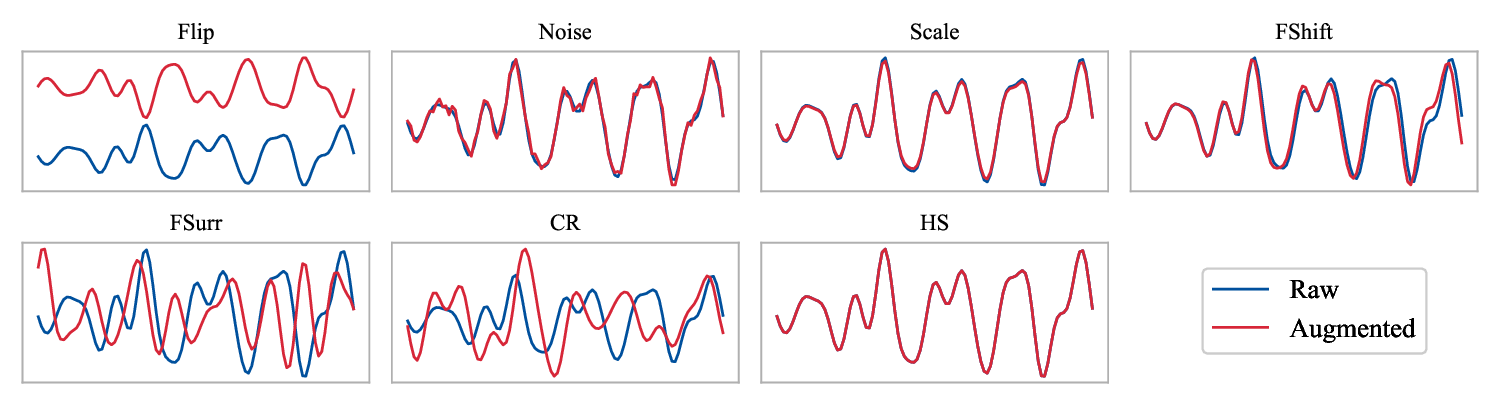}
\caption{Visualizations of EEG trials before (blue lines) and after (red lines) seven data augmentation approaches.} \label{fig:AugExample}
\end{figure*}

Operations of data augmentation approaches are detailed in Table \ref{tab:da_info}. Visualizations of EEG trials before and after data augmentation are shown in Figure~\ref{fig:AugExample}.

\begin{table}[htpb]
\small
\centering \setlength{\tabcolsep}{1mm}
\caption{Operations of data augmentation strategies.}  \label{tab:da_info}
\begin{tabular}{c|c|c|c}
\toprule
Type & Strategy & Formulation & Parameter\\
\midrule
\multirow{3}{*}{Time}
& Flip & $\tilde{X} = \max(X) - X$ & - \\
& Noise & $\tilde{X} = X + rand \ast std(X) / C_{\text{noise}}$ & $C_{\text{noise}}=2$ \\
& Scale & $\tilde{X} = X \ast (1 \pm C_{\text{scale}})$ & $C_{\text{scale}}=0.05$ \\
\midrule
\multirow{2}{*}{Frequency}
& FShift & $\tilde{X} = F_{\text{shift}}(X, \pm C_{\text{shift}})$ & $C_{\text{shift}}=0.2$ \\
& FSurr & $\tilde{X} = F_{\text{surr}}(X, C_{\text{surr}})$ & $C_{\text{surr}}=0.4$ \\
\midrule
\multirow{2}{*}{Space}
& CR & $X_L \leftrightarrow X_R, Y = 1-Y$ & - \\
& HS & $\tilde{X} = X^i_L\oplus X^{j}_R, i \neq j$ & - \\
\bottomrule
\end{tabular}
\end{table}

\subsection{Cross-View Contrasting}

The CVC module is designed to enhance the consistency of features across different augmented views. The raw trial is treated as the anchor, and paired with its augmented counterparts in time, space, and frequency domains, resulting in three positive pairs: raw-time, raw-space, and raw-frequency. Negative samples are composed of all other trials and their augmentations.

\begin{figure}[htpb]\centering
\includegraphics[width=.9\linewidth,clip]{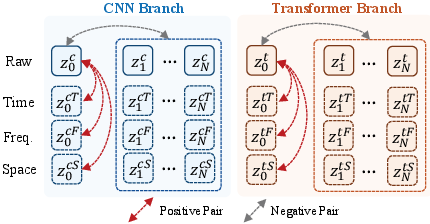}
\caption{Illustration of positive and negative pairs in the CVC module.}
\label{fig:cvc}
\end{figure}

Given a mini-batch of $N$ trials, a total of $3N$ augmented trials are generated. For each trial, three positive pairs and $4(N-1)$ negative pairs are constructed in each branch (CNN or Transformer), resulting in $6$ positive pairs and $8(N-1)$ negative pairs in total, as illustrated in Figure~\ref{fig:cvc}.

We adopt the NT-Xent loss \cite{simclr2020} as the contrastive objective, aiming to maximize the similarity between positive pairs while minimizing it for negative pairs. The contrastive distance for the extracted feature $\mathbf{z}_i$ is computed as:
\begin{align}
d(\mathbf{z}_{i}, \mathbf{z}_{i}^{v}) = -\log \frac{
\exp \left( \mathrm{sim}\left(\mathbf{z}_{i}, \mathbf{z}_{i}^{v}\right) / \tau \right)
}{
\sum\limits_{j=1}^{N} \mathbb{I}_{[i \neq j]}
\exp \left( \mathrm{sim} \left(\mathbf{z}_{i}, \mathbf{z}_{j}^{v}\right) / \tau \right)
}, \label{eq:TFC}
\end{align}
where $v \in \{T, S, F\}$ denotes the view type, $\mathrm{sim}(\cdot,\cdot)$ denotes cosine similarity, $N$ is the number of samples in a mini-batch, $\tau$ is a temperature hyperparameter, and $\mathbb{I}_{[i \neq j]}$ is an indicator function that equals 1 if $i \neq j$, and 0 otherwise.

The overall CVC loss is formulated as:
\begin{align}
\mathcal{L}_\mathrm{CVC} = \frac{1}{N V} \sum_{i=1}^{N} \sum_{v \in \{T, S, F\}} \left(
d(\mathbf{z}_{i}^{c}, \mathbf{z}_{i}^{cv}) +
d(\mathbf{z}_{i}^{t}, \mathbf{z}_{i}^{tv})
\right), \label{eq:cvc_loss}
\end{align}
where $c$ and $t$ denote the CNN and Transformer branches, respectively.

\subsection{Cross-Model Contrasting}

The CMC module is introduced to align the representations learned from the two network branches. Positive pairs are features extracted from the same trial by different branches, while negative pairs are features from different trials, including both raw and augmented samples.

\begin{figure}[htpb]\centering
\includegraphics[width=.9\linewidth,clip]{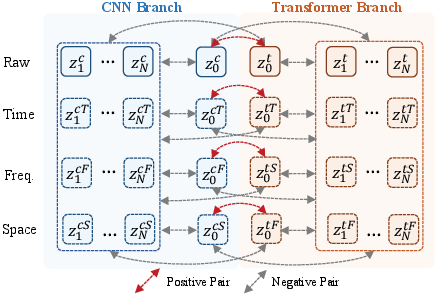}
\caption{Illustration of positive and negative pairs in the CMC module.}
\label{fig:cmc}
\end{figure}

For each sample, one positive pair and $4(N-1)$ negative pairs are constructed per view, yielding four positive pairs and $16(N-1)$ negative pairs in total, as shown in Figure~\ref{fig:cmc}.

The NT-Xent loss is also used to compute the contrastive distance between CNN and Transformer features:
\begin{align}
d(\mathbf{z}_i^{cv}, \mathbf{z}_i^{tv}) = -\log \frac{
\exp \left( \mathrm{sim}\left(\mathbf{z}_{i}^{cv}, \mathbf{z}_{i}^{tv}\right) / \tau \right)
}{
\sum\limits_{j=1}^{N} \mathbb{I}_{[i \neq j]}
\exp \left( \mathrm{sim} \left(\mathbf{z}_{i}^{cv}, \mathbf{z}_{j}^{tv}\right) / \tau \right)
}, \label{eq:TSFC}
\end{align}
where $\mathbf{z}_i^{cv}$ and $\mathbf{z}_i^{tv}$ represent CNN and Transformer features with the same feature dimension of sample $i$ under view $v$.

The final CMC loss is given by:
\begin{align}
\mathcal{L}_\mathrm{CMC} = \frac{1}{N V} \sum_{i=1}^{N} \left(
d(\mathbf{z}_i^{c}, \mathbf{z}_i^{t}) +
\sum_{v \in \{T, S, F\}} d(\mathbf{z}_i^{cv}, \mathbf{z}_i^{tv})
\right), \label{eq:cmc_loss}
\end{align}
where $d(\mathbf{z}_i^c, \mathbf{z}_i^t)$ denotes the contrastive distance between features of the raw trial.

\subsection{MVCNet Objective}

The classification loss $\mathcal{L}_\mathrm{CLS}$ is computed by summing the cross-entropy losses over the original trial and its augmented views. Specifically, for each sample:
\begin{align}
\mathcal{L}_\mathrm{CLS} = \sum_{i=1}^{N} \left(
\mathrm{CE}(\hat{y}_i, y_i) +
\sum_{v \in \{T, S, F\}} \mathrm{CE}(\hat{y}_i^v, y_i)
\right), \label{eq:cls_loss}
\end{align}
where $\mathrm{CE}(\cdot,\cdot)$ denotes the standard cross-entropy loss, and $v$ indexes the three augmented views.

The overall optimization objective of MVCNet combines the classification loss and two contrastive regularization terms as:
\begin{align}
\mathcal{L}_\mathrm{all} =
\mathcal{L}_\mathrm{CLS} +
\lambda \cdot \mathcal{L}_\mathrm{CVC} +
\gamma \cdot \mathcal{L}_\mathrm{CMC}, \label{eq:TSF_loss}
\end{align}
where $\lambda$ and $\gamma$ are trade-off hyperparameters balancing the contributions of the contrastive losses.

The pseudo-code of MVCNet is given in Algorithm~\ref{alg:MVCNet}.

\begin{algorithm}[htb]
\caption{Multi-View Contrastive Network (MVCNet)}
\label{alg:MVCNet}
\begin{algorithmic}[1]
\REQUIRE
Labeled training data $\{(\mathbf{x}_i, y_i)\}_{i=1}^{n_s}$;\\
Unlabeled test data $\{\mathbf{x}_j^t\}_{j=1}^{n_t}$;\\
CNN branch $B_c$, Transformer branch $B_t$, Classifier $F$;\\
Data augmentation functions $A_T$, $A_S$, $A_F$ with views $v \in \{T, S, F\}$;\\
Batch size $N$; temperature $\tau$; loss weights $\lambda$, $\gamma$.
\ENSURE
Predicted labels $\{\hat{y}_j^t\}_{j=1}^{n_t}$ for test data $\{\mathbf{x}_j^t\}_{j=1}^{n_t}$.
\vspace{0.5em}
\WHILE{training not converged}
    \STATE Sample a batch $\{(\mathbf{x}_i, y_i)\}_{i=1}^{N}$ from training data;
	\STATE Generate augmented views:
	\STATE \quad $\mathbf{x}_i^T = A_T(\mathbf{x}_i)$
	\STATE \quad $\mathbf{x}_i^S = A_S(\mathbf{x}_i)$
	\STATE \quad $\mathbf{x}_i^F = A_F(\mathbf{x}_i)$

	\STATE Extract CNN branch features:
	\STATE \quad $\mathbf{z}_i^{c} = B_c(\mathbf{x}_i)$
	\STATE \quad $\mathbf{z}_i^{cv} = B_c(\mathbf{x}_i^v)$

	\STATE Extract Transformer branch features:
	\STATE \quad $\mathbf{z}_i^{t} = B_t(\mathbf{x}_i)$
	\STATE \quad $\mathbf{z}_i^{tv} = B_t(\mathbf{x}_i^v)$
    \STATE Fuse features from both branches for all views $v$:
    \STATE \quad $\mathbf{z}_i = \texttt{Concat}(\mathbf{z}_i^c,\ \mathbf{z}_i^t)$
    \STATE \quad $\mathbf{z}_i^v = \texttt{Concat}(\mathbf{z}_i^{cv},\ \mathbf{z}_i^{tv})$
    \STATE Predict category:
    \STATE \quad $\hat{y}_i = F(\mathbf{z}_i)$
    \STATE \quad $\hat{y}_i^v = F(\mathbf{z}_i^v)$
    \STATE Compute cross-view contrastive loss $\mathcal{L}_\mathrm{CVC}$ across views $\mathbf{z}_i, \mathbf{z}_i^{v}$ for each branch by Eq.~(\ref{eq:TFC})--(\ref{eq:cvc_loss});
    \STATE Compute cross-model contrastive loss $\mathcal{L}_\mathrm{CMC}$ between $\mathbf{z}_i^c, \mathbf{z}_i^{cv}$ and $\mathbf{z}_i^t, \mathbf{z}_i^{tv}$ by Eq.~(\ref{eq:TSFC})--(\ref{eq:cmc_loss});
    \STATE Compute classification loss $\mathcal{L}_\mathrm{CLS}$ by Eq.~(\ref{eq:cls_loss}) between $\hat{y}_i, \hat{y}_i^v$  and $y_i$;
    \STATE Compute total loss: $\mathcal{L}_{\mathrm{all}} = \mathcal{L}_{\mathrm{CLS}} + \lambda \mathcal{L}_{\mathrm{CVC}} + \gamma \mathcal{L}_{\mathrm{CMC}}$;
    \STATE Update $B_c$, $B_t$, and $F$ by minimizing $\mathcal{L}_{\mathrm{all}}$;
\ENDWHILE
\vspace{0.5em}
\STATE \textbf{Return} $\{\mathbf{y}_j^t\}_{j=1}^{n_t} = F(\texttt{Concat}(B_c(\mathbf{x}_j^t),\ B_t(\mathbf{x}_j^t)))$.
\end{algorithmic}
\end{algorithm}

\section{Experiments and Results}\label{sect:er}
This section presents the datasets, experiments and analyses. Code is available on GitHub\footnote{https://github.com/wzwvv/MVCNet}.

\subsection{Datasets}
Four EEG-based MI benchmark datasets, namely BNCI2014001 \cite{tangermann2014001}, Zhou2016 \cite{Zhou2016}, BNCI2014002 \cite{Steyrl2016BNCI2014002}, and BNCI2015001 \cite{Faller2012BNCI2015001} datasets from the Mother Of All BCI Benchmarks (MOABB) \cite{Sylvain2024} were used. An additional Blankertz2007 \cite{Blankertz2007MI1} dataset from BCI Competition IV-1 was also used. Their characteristics are summarized in Table \ref{tab:dataset_info}.
\begin{table*}[htpb]  \centering \setlength{\tabcolsep}{2mm}
\caption{Summary of the five MI datasets.}
\small
\label{tab:dataset_info}
\begin{tabular}{c|c|c|c|c|c|c}
\toprule
\multirow{2}{*}{Dataset} & Number of & Number of & Sampling & Trial Length & Number of & \multirow{2}{*}{Task Types} \\
 & Subjects & EEG Channels & Rate (Hz) & (seconds) & Total Trials & \\
\midrule
BNCI2014001 & 9 & 22 & 250 & 4 & 1296 & left/right hand \\
Zhou2016 & 4 & 14 & 250 & 5 & 409 & left/right hand\\
Blankertz2007 & 7 & 59 & 250 & 3 & 1,400 & left/right hand or left hand/right foot \\
BNCI2014002 & 14 & 15 & 512 & 5 & 1,400 & right hand/feet \\
BNCI2015001 & 12 & 13 & 512 & 5 & 2,400 & right hand/feet \\
\bottomrule
\end{tabular}
\end{table*}

For the BNCI2014001, Zhou2016, BNCI2014002, and BNCI2015001 datasets, the standard preprocessing steps in MOABB, including notch filtering, band-pass filtering, etc., were used to ensure the reproducibility. For the Blankertz2007 dataset, the EEG trials were first band-pass filtered between 8 and 30 Hz. Trials between [0.5, 3.5] seconds after the cue onset were used and then downsampled to 250 Hz.

Euclidean Alignment (EA) \cite{He2020EA}, an effective unsupervised EEG data alignment approach \cite{Wu2022NN}, was utilized after pre-processing. In the cross-subject scenario, the EA reference matrix of the target subject was updated as new test trials arrived on-the-fly, as in \cite{Li2024T-TIME}.

\subsection{Compared Approaches}
Nine MI decoding neural networks were reproduced and compared with the proposed MVCNet.
\begin{itemize}
\item EEGNet \cite{Lawhern2018EEGNet} is a lightweight CNN architecture specifically designed for EEG classification. It comprises two convolutional blocks and a classification head. The model begins with a temporal convolution to capture frequency-specific patterns, followed by a depthwise convolution to extract spatial features. A separable convolution, combined with a pointwise convolution, further enhances feature extraction.
\item DCNN \cite{deepshallow2017} is a deep CNN-based architecture with a larger parameter count. It consists of four convolutional max-pooling blocks, followed by a fully connected softmax layer for classification.
\item SCNN \cite{deepshallow2017} is a simplified variant of DCNN, inspired by the filter bank common spatial pattern. It employs two layers dedicated to temporal convolution and spatial filtering to enhance feature extraction efficiency.
\item FBCNet \cite{mane2021fbcnet} integrates a spatial convolutional layer and a temporal variance layer to extract spectral-spatial features from multi-band EEG trials. A fully connected layer is then applied for classification.
\item ADFCNN \cite{tao2023adfcnn} is a dual-scale CNN architecture that employs temporal convolution to capture frequency-domain features and spatial convolution in a separable manner to extract global spatial information. The extracted features from the two branches are fused using an attention-based feature fusion module, followed by a fully connected classification layer.
\item SlimSeiz \cite{lu2024slimseiz} employs multiple stacked 1D convolutional layers to extract temporal features at varying resolutions, followed by a lightweight Mamba block to model long-range dependencies. By leveraging the Mamba architecture, SlimSeiz achieves effective sequence modeling with less parameter overhead.
\item CTNet \cite{zhao2024ctnet} introduces a sequential architecture that integrates a CNN block, similar to EEGNet, with a Transformer block. The convolutional layers capture local spatio-temporal features, while the Transformer enhances global temporal modeling.
\item EEG Conformer \cite{song2022eeg} comprises a convolutional module, a self-attention module, and a classifier module. Temporal and spatial convolutional layers are applied along the temporal and channel dimensions, followed by an average pooling layer. The Transformer module further captures long-term temporal dependencies. Finally, multiple fully connected layers are employed for classification.
\item IFNet \cite{wang2023ifnet}, akin to FBCNet, first decomposes EEG signals into multiple frequency bands (e.g., 4-16 Hz and 16-40 Hz). A combination of 1D spatial convolution and 1D temporal convolution is then utilized to extract spectral-spatial representations, followed by a fully connected layer for final classification.
\end{itemize}

To further assess the efficacy of data augmentation, we compared the proposed MVCNet with seven data augmentation approaches introduced in Table \ref{tab:da_info}:
\begin{itemize}
\item Baseline, trained using standard cross-entropy loss without data augmentation.
\item Seven EEG data augmentation strategies, evaluated under a supervised learning framework. Among them, CR, HS, and Flip do not require hyperparameters, while Noise, Scale, Freq, and Surr involve hyperparameters, whose values are determined based on \cite{Zhang2022MSDT,pei2021hs}.
\end{itemize}

\subsection{Implementation}
Three evaluation scenarios were considered to assess the generalization ability of MVCNet: chronological order (CO), cross-validation (CV), and leave-one-subject-out (LOSO), including within/cross-subject settings.
\begin{itemize}
\item CO: EEG trials were partitioned strictly based on temporal sequence, with the first 80\% used for training and the remaining 20\% for testing. This setting follows a within-subject evaluation paradigm.
\item CV: A 5-fold cross-validation strategy was employed, where four folds were used for training and one for testing. The data partitions were structured chronologically while maintaining class-balance, following \cite{mane2021fbcnet}. CV is also a within-subject setting.
\item LOSO: EEG trials from one subject were reserved for testing, while all other subjects' trials were combined for training. LOSO is a cross-subject setting.
\end{itemize}

All experiments were repeated five times with random seed list $[1, 2, 3, 4, 5]$, and the average results were reported. For all datasets, models were trained for 100 epochs using the Adam optimizer with learning rate $10^{-3}$. The temperature parameter $\tau$ was set to 0.2, following \cite{chen2020simclrv2}. $\lambda$ and $\gamma$ were fixed at 0.1 for all datasets. Batch sizes were set to 32 for baseline models, and 64 for data augmentation approaches and MVCNet, depending on the size of the training data. For MVCNet, different combinations of augmentations were applied based on the MI task: Scale, FShift, and CR were used for the BNCI2014001, Zhou2016, and Blankertz2007 datasets, while Flip, FShift, and HS were used for the BNCI2014002 and BNCI2015001 datasets. For all baseline models, architectural hyperparameters (e.g., kernel sizes, number of layers) were set according to their original papers to ensure fair comparison and reproducibility.

\subsection{Main Results}
Table~\ref{tab:main_results} reports the average classification accuracies (\%) of proposed MVCNet and nine baseline models across five public MI datasets under three scenarios. Observe that:
\begin{itemize}
\item CV consistently outperforms CO across all datasets and models. For example, EEGNet improves from 73.92\% (CO) to 76.66\% (CV), and IFNet from 81.15\% to 84.90\%. This trend highlights the critical role of data partitioning: In the CO setting, temporal distribution shifts are introduced due to user fatigue, attentional drift, or electrode impedance changes, all of which can lead to degraded generalization. In contrast, each fold in the CV protocol contains a temporally contiguous segment of the recording session, ensuring realistic testing while maintaining class balance. As a result, CV yields improved generalization without violating realistic deployment assumptions, allowing the model learning a wider range of temporal patterns during training, while still preserving temporal independence between training and test sets. This underscores the importance of fair and appropriate partitioning when benchmarking within-subject EEG decoding models.
\item Among the nine baselines, IFNet achieves the highest average accuracy in CO (81.15\%) and CV (84.90\%), benefiting from frequency-aware design. EEG Conformer ranks second with 78.71\% and 83.91\%. Under the LOSO setting, IFNet remains best (76.16\%), followed by DCNN (75.46\%), indicating that deeper CNNs better capture cross-subject variance.
\item The proposed MVCNet consistently outperforms all baselines in all scenarios and datasets. It surpasses the second best IFNet by +3.11\%, +2.71\%, and +2.56\%, respectively. These results validate the effectiveness of our dual-branch design and multi-view contrastive framework.
\item MVCNet maintains strong performance on the challenging LOSO setting. This confirms its ability to learn consistent and generalizable EEG representations, making it a robust solution for practical EEG decoding.
\end{itemize}

\begin{table*}[htpb]
\centering
\footnotesize
\setlength{\tabcolsep}{0.6mm}
\renewcommand\arraystretch{1.2}
\caption{Average classification accuracies (\%) of MVCNet and nine baseline models on five MI datasets in CO, CV, and LOSO scenarios. The best average performance of each dataset is marked in bold, and the second best by an underline.}
\label{tab:main_results}
\begin{tabular}{c|c|cccccccccc}
\toprule
Scenario & Dataset & EEGNet & SCNN & DCNN & FBCNet & ADFCNN & SlimSeiz & CTNet & EEG Conformer & IFNet & MVCNet\\
\midrule
\multirow{6.5}{*}{CO}
& BNCI2014001 & 69.05$_{\pm1.00}$ & 73.57$_{\pm2.36}$ & 59.29$_{\pm1.64}$ & 68.97$_{\pm1.26}$ & 73.73$_{\pm2.26}$ & 68.89$_{\pm2.05}$ & 73.49$_{\pm2.05}$ & \underline{78.57}$_{\pm0.66}$ & 77.94$_{\pm0.93}$ & \textbf{83.17}$_{\pm0.74}$ \\
& Zhou2016 & 80.13$_{\pm3.35}$ & 75.03$_{\pm6.15}$ & 78.03$_{\pm2.37}$ & 63.33$_{\pm2.29}$ & 71.42$_{\pm1.95}$ & 72.01$_{\pm3.98}$ & 76.81$_{\pm5.05}$ & 73.87$_{\pm4.51}$ & \underline{81.70}$_{\pm2.08}$ & \textbf{84.11}$_{\pm2.68}$ \\
& Blankertz2007 & 78.79$_{\pm2.78}$ & 76.71$_{\pm2.09}$ & 70.00$_{\pm4.12}$ & 75.93$_{\pm1.31}$ & 76.07$_{\pm1.41}$ & 65.64$_{\pm1.18}$ & 79.00$_{\pm3.38}$ & 82.29$_{\pm2.62}$ & \underline{84.00}$_{\pm0.57}$ & \textbf{87.07}$_{\pm0.52}$ \\
& BNCI2014002 & 66.07$_{\pm2.76}$ & \underline{79.07}$_{\pm1.96}$ & 64.07$_{\pm2.70}$ & 69.50$_{\pm0.95}$ & 73.00$_{\pm1.95}$ & 78.71$_{\pm0.53}$ & 71.00$_{\pm1.80}$ & 76.21$_{\pm1.46}$ & 78.29$_{\pm1.68}$ & \textbf{81.29}$_{\pm2.31}$ \\
& BNCI2015001 & 75.58$_{\pm1.69}$ & 83.71$_{\pm1.34}$ & 71.08$_{\pm1.82}$ & 74.92$_{\pm0.97}$ & 78.75$_{\pm0.62}$ & 81.71$_{\pm0.81}$ & 78.21$_{\pm1.09}$ & 82.63$_{\pm0.54}$ & \underline{83.83}$_{\pm0.90}$ & \textbf{85.67}$_{\pm0.55}$ \\
\cmidrule{2-12}
& Average & 73.92 & 77.62 & 68.49 & 70.53 & 74.59 & 73.39 & 75.70 & 78.71 & \underline{81.15} & \textbf{84.26}\\
\midrule
\multirow{6.5}{*}{CV}
& BNCI2014001 & 71.86$_{\pm0.96}$ & 78.22$_{\pm0.19}$ & 61.59$_{\pm1.08}$ & 74.74$_{\pm0.67}$ & 77.27$_{\pm0.60}$ & 75.56$_{\pm1.48}$ & 75.97$_{\pm0.76}$ & \underline{83.62}$_{\pm0.99}$ & 82.62$_{\pm0.91}$ & \textbf{84.49}$_{\pm0.61}$ \\
& Zhou2016 & 85.30$_{\pm2.12}$ & 82.90$_{\pm0.90}$ & 79.68$_{\pm1.13}$ & 80.01$_{\pm0.50}$ & 84.95$_{\pm1.51}$ & 85.60$_{\pm1.36}$ & 86.38$_{\pm1.45}$ & 85.67$_{\pm1.52}$ & \underline{87.78}$_{\pm1.01}$ & \textbf{91.72}$_{\pm0.27}$ \\
& Blankertz2007 & 80.21$_{\pm0.77}$ & 81.11$_{\pm0.72}$ & 69.83$_{\pm1.51}$ & 84.14$_{\pm0.67}$ & 81.10$_{\pm1.75}$ & 75.37$_{\pm1.14}$ & 83.64$_{\pm1.04}$ & 87.43$_{\pm0.33}$ & \underline{88.23}$_{\pm0.50}$ & \textbf{90.81}$_{\pm0.64}$ \\
& BNCI2014002 & 68.90$_{\pm1.37}$ & 81.24$_{\pm0.50}$ & 67.17$_{\pm0.31}$ & 74.53$_{\pm0.68}$ & 74.63$_{\pm0.75}$ & 77.93$_{\pm0.85}$ & 75.50$_{\pm0.78}$ & 79.30$_{\pm0.33}$ & \underline{80.21}$_{\pm0.52}$ & \textbf{83.34}$_{\pm0.55}$ \\
& BNCI2015001 & 77.04$_{\pm0.72}$ & 85.58$_{\pm0.42}$ & 76.89$_{\pm1.05}$ & 79.57$_{\pm0.27}$ & 80.69$_{\pm0.40}$ & 81.68$_{\pm0.50}$ & 80.94$_{\pm0.42}$ & 83.53$_{\pm0.71}$ & \underline{85.67}$_{\pm0.56}$ & \textbf{87.68}$_{\pm0.24}$ \\
\cmidrule{2-12}
& Average & 76.66 & 81.81 & 71.03 & 78.60 & 79.73 & 79.23 & 80.49 & 83.91 & \underline{84.90} & \textbf{87.61}\\
\midrule
\multirow{6.5}{*}{LOSO}
&BNCI2014001 & 73.64$_{\pm1.14}$ & 72.22$_{\pm1.03}$ & 73.21$_{\pm1.79}$ & 72.56$_{\pm0.96}$ & 71.76$_{\pm0.75}$ & 69.62$_{\pm1.32}$ & 73.40$_{\pm1.22}$ & 73.07$_{\pm2.01}$ & \underline{74.52}$_{\pm0.75}$ & \textbf{76.02}$_{\pm0.57}$ \\
& Zhou2016 & 83.22$_{\pm1.83}$ & 82.10$_{\pm0.67}$ & 83.84$_{\pm1.18}$ & 82.07$_{\pm0.91}$ & 82.09$_{\pm1.31}$ & 84.06$_{\pm1.58}$ & 83.88$_{\pm1.03}$ & 82.43$_{\pm1.48}$ & \underline{86.21}$_{\pm0.99}$ & \textbf{86.96}$_{\pm1.07}$ \\
& Blankertz2007 & 71.10$_{\pm0.83}$ & 70.64$_{\pm0.58}$ & 72.19$_{\pm0.89}$ & \underline{76.23}$_{\pm1.41}$ & 70.59$_{\pm1.69}$ & 73.21$_{\pm0.82}$ & 69.50$_{\pm1.79}$ & 74.41$_{\pm1.13}$ & 73.43$_{\pm1.06}$ & \textbf{79.56}$_{\pm1.05}$ \\
& BNCI2014002 & 72.86$_{\pm0.38}$ & 70.57$_{\pm1.35}$ & \underline{74.34}$_{\pm0.81}$ & 71.31$_{\pm0.82}$ & 72.67$_{\pm0.44}$ & 73.01$_{\pm0.60}$ & 74.14$_{\pm0.79}$ & 72.84$_{\pm1.36}$ & \underline{73.90}$_{\pm0.65}$ & \textbf{75.24}$_{\pm0.51}$ \\
& BNCI2015001 & 71.89$_{\pm0.70}$ & 69.30$_{\pm0.88}$ & \underline{73.73}$_{\pm0.32}$ & 67.61$_{\pm0.99}$ & 71.86$_{\pm0.89}$ & 70.74$_{\pm0.82}$ & 72.33$_{\pm0.74}$ & 70.88$_{\pm0.68}$ & 72.73$_{\pm1.17}$ & \textbf{75.85}$_{\pm0.45}$ \\
\cmidrule{2-12}
& Average & 74.54 & 72.97 & 75.46 & 73.96 & 73.79 & 74.13 & 74.65 & 74.73 & \underline{76.16} & \textbf{78.72}\\
\bottomrule
\end{tabular}
\end{table*}

\subsection{Comparison with Data Augmentation Strategies}
To further evaluate the effectiveness of MVCNet, we compare it with seven representative data augmentation strategies under the LOSO setting. Specifically, we replace the CNN backbone in our framework with four current architectures: SCNN, DCNN, EEGNet, and IFNet, and apply each augmentation individually. The classification results across four MI datasets are presented in Table~\ref{tab:basic_results}. Observe that:
\begin{itemize}
\item Among the four CNN backbones, IFNet consistently achieves the highest average accuracy, followed by EEGNet.
\item The performance of individual augmentation techniques varies across datasets and models. For example, Surr significantly degrades performance on SCNN, while HS shows poor results on DCNN. Although CR achieves good performance on Zhou2016 and Blankertz2007, it is less effective on BNCI2014002 and BNCI2015001, due to the asymmetric motor imagery classes (e.g., right hand vs. both feet).
\item MVCNet outperforms most augmentation strategies across all CNN backbones, obtaining the best or second-best performance. This indicates that MVCNet integrates more informative transformations and learns more robust and discriminative EEG representations under the challenging cross-subject scenario.
\end{itemize}

\begin{table*}[h]
\centering \small
\setlength{\tabcolsep}{0.8mm}
\renewcommand\arraystretch{1.2}
\caption{Average classification accuracies (\%) with four networks under LOSO setting. The best average performance of each network is marked in bold, and the second best by an underline.}  \label{tab:basic_results}
    \begin{tabular}{c|c|ccccccccc}   \toprule
Backbone & Dataset & Baseline & Flip & Noise & Scale & Shift & Surr & CR & HS & MVCNet(Ours) \\
\midrule
\multirow{5.5}{*}{SCNN}
&Zhou2016 & 81.97$_{\pm1.63}$ & 81.82$_{\pm0.74}$ & 82.62$_{\pm1.20}$ & \underline{83.67}$_{\pm0.91}$ & 81.67$_{\pm0.61}$ & 63.04$_{\pm1.20}$ & 81.87$_{\pm0.78}$ & 81.57$_{\pm0.58}$ & \textbf{84.06}$_{\pm1.02}$ \\
& Blankertz2007 & 70.04$_{\pm0.79}$ & \underline{74.10}$_{\pm0.90}$ & 73.59$_{\pm0.90}$ & 71.96$_{\pm0.87}$ & 73.57$_{\pm0.44}$ & 61.51$_{\pm0.88}$ & 73.89$_{\pm0.66}$ & 73.38$_{\pm0.63}$ & \textbf{75.09}$_{\pm0.63}$ \\
& BNCI2014002 & 68.13$_{\pm0.74}$ & 72.10$_{\pm0.37}$ & \textbf{72.97}$_{\pm0.41}$ & 72.36$_{\pm0.74}$ & 72.28$_{\pm0.70}$ & 66.74$_{\pm1.14}$ & 72.43$_{\pm0.45}$ & 71.72$_{\pm0.67}$ & \underline{72.67}$_{\pm1.09}$ \\
& BNCI2015001 & 69.71$_{\pm0.67}$ & \underline{69.80}$_{\pm0.89}$ & 69.31$_{\pm1.00}$ & 68.98$_{\pm0.82}$ & 69.35$_{\pm0.90}$ & 68.60$_{\pm0.65}$ & 69.50$_{\pm0.73}$ & 68.32$_{\pm0.96}$ & \textbf{71.74}$_{\pm0.98}$ \\
\cmidrule{2-11}
& Average & 72.46 & 74.46 & \underline{74.62} & 74.24 & 74.22 & 64.97 & 74.42 & 73.75 & \textbf{75.89} \\
\midrule
\multirow{5.5}{*}{DCNN}
& Zhou2016 & 82.91$_{\pm0.85}$ & 84.01$_{\pm1.21}$ & 83.27$_{\pm0.44}$ & 83.22$_{\pm1.30}$ & 83.69$_{\pm1.13}$ & 78.76$_{\pm1.73}$ & \underline{84.22}$_{\pm1.26}$ & 53.88$_{\pm1.43}$ & \textbf{85.19}$_{\pm1.26}$ \\
& Blankertz2007 & 71.44$_{\pm0.78}$ & 71.31$_{\pm0.76}$ & 71.06$_{\pm1.32}$ & 71.80$_{\pm0.46}$ & 70.87$_{\pm0.52}$ & 67.39$_{\pm0.76}$ & \underline{72.10}$_{\pm1.50}$ & 50.72$_{\pm0.42}$ & \textbf{74.03}$_{\pm1.43}$ \\
& BNCI2014002 & 69.74$_{\pm0.94}$ & 74.76$_{\pm0.97}$ & 74.74$_{\pm1.12}$ & \underline{74.98}$_{\pm0.80}$ & \textbf{75.14}$_{\pm0.87}$ & 72.23$_{\pm0.46}$ & 74.35$_{\pm1.20}$ & 54.41$_{\pm2.60}$ & 74.36$_{\pm0.88}$ \\
& BNCI2015001 & 70.42$_{\pm0.68}$ & 73.65$_{\pm0.82}$ & \underline{74.20}$_{\pm0.41}$ & 74.32$_{\pm0.86}$ & 74.12$_{\pm0.75}$ & 73.86$_{\pm0.62}$ & 73.93$_{\pm0.35}$ & 61.78$_{\pm2.59}$ & \textbf{76.11}$_{\pm0.61}$ \\
\cmidrule{2-11}
& Average & 73.63 & 75.93 & 75.82 & 76.08 & 75.96 & 73.06 & \underline{76.15} & 55.20 & \textbf{77.42} \\
\midrule
\multirow{5.5}{*}{EEGNet}
& Zhou2016 & 83.22$_{\pm1.73}$ & 81.19$_{\pm2.39}$ & 84.16$_{\pm0.96}$ & 83.99$_{\pm0.83}$ & 82.94$_{\pm1.99}$ & 83.82$_{\pm1.18}$ & \underline{84.82}$_{\pm1.36}$ & 80.18$_{\pm2.52}$ & \textbf{87.20}$_{\pm1.21}$ \\
& Blankertz2007 & 71.17$_{\pm0.87}$ & 69.86$_{\pm1.49}$ & 71.87$_{\pm0.55}$ & 71.70$_{\pm0.73}$ & 70.96$_{\pm0.82}$ & 69.82$_{\pm0.56}$ & \underline{74.97}$_{\pm0.96}$ & 68.31$_{\pm2.68}$ & \textbf{76.56}$_{\pm1.23}$ \\
& BNCI2014002 & 72.86$_{\pm0.38}$ & \underline{73.75}$_{\pm1.31}$ & 72.49$_{\pm0.69}$ & 72.59$_{\pm0.68}$ & 73.51$_{\pm1.07}$ & 72.21$_{\pm0.97}$ & 72.43$_{\pm0.70}$ & 69.99$_{\pm2.37}$ & \textbf{75.24}$_{\pm0.87}$ \\
& BNCI2015001 & 71.89$_{\pm0.70}$ & 71.96$_{\pm1.11}$ & 72.28$_{\pm1.02}$ & 71.73$_{\pm1.30}$ & 73.11$_{\pm1.31}$ & \underline{73.21}$_{\pm1.20}$ & 72.21$_{\pm0.93}$ & 70.91$_{\pm2.17}$ & \textbf{75.93}$_{\pm0.94}$ \\
\cmidrule{2-11}
& Average & 74.79 & 74.19 & 75.20 & 75.00 & 75.13 & 74.77 & \underline{76.11} & 72.35 & \textbf{78.65} \\
\midrule
\multirow{5.5}{*}{IFNet}
& Zhou2016 & \underline{86.21}$_{\pm0.99}$ & 85.71$_{\pm0.41}$ & 85.66$_{\pm0.57}$ & 86.17$_{\pm0.49}$ & 85.91$_{\pm0.69}$ & 79.46$_{\pm1.33}$ & 86.05$_{\pm0.62}$ & 85.38$_{\pm0.95}$ & \textbf{87.36}$_{\pm1.87}$ \\
& Blankertz2007 & 73.43$_{\pm1.06}$ & 76.16$_{\pm0.43}$ & 76.16$_{\pm0.37}$ & 75.83$_{\pm0.52}$ & 76.00$_{\pm0.74}$ & 75.09$_{\pm0.86}$ & \textbf{79.36}$_{\pm1.33}$ & 72.99$_{\pm1.25}$ & \underline{77.53}$_{\pm0.57}$ \\
& BNCI2014002 & 73.90$_{\pm0.65}$ & 75.33$_{\pm1.02}$ & 75.93$_{\pm0.69}$ & \underline{75.96}$_{\pm0.57}$ & 75.83$_{\pm1.13}$ & 75.14$_{\pm0.71}$ & 71.44$_{\pm0.45}$ & 75.89$_{\pm0.64}$ & \textbf{76.37}$_{\pm0.51}$ \\
& BNCI2015001 & 72.73$_{\pm1.17}$ & \underline{72.96}$_{\pm0.60}$ & 72.43$_{\pm0.85}$ & 72.54$_{\pm0.73}$ & 73.13$_{\pm0.53}$ & 72.30$_{\pm0.43}$ & 70.83$_{\pm0.76}$ & 71.32$_{\pm1.87}$ & \textbf{76.23}$_{\pm0.24}$ \\
\cmidrule{2-11}
& Average & 76.57 & 77.54 & 77.54 & 77.62 & \underline{77.72} & 75.50 & 76.92 & 76.39 & \textbf{79.37} \\
\bottomrule
\end{tabular}
\end{table*}

\subsection{Effect of Multi-View Augmentation}
\begin{figure*}[h]\centering
\includegraphics[width=\linewidth,clip]{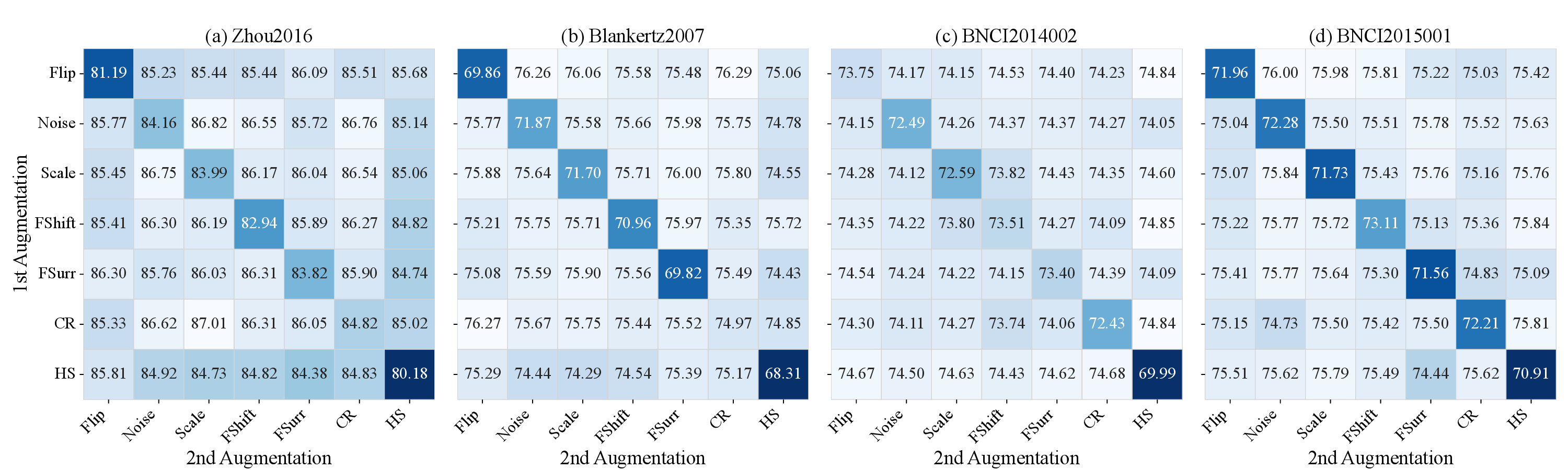}
\caption{Classification accuracies (\%) for all pairwise combinations of seven data augmentation approaches under the LOSO setting using EEGNet as the backbone. (a) Zhou2016, (b) Blankertz2007, (c) BNCI2014002, and (d) BNCI2015001.} \label{fig:augzoo}
\end{figure*}

To investigate the impact of combining different augmentation views, we evaluated all pairwise combinations among the seven augmentation strategies using EEGNet as the backbone. As shown in Figure~\ref{fig:augzoo}, combinations of two views generally yield better performance than individual augmentations, demonstrating the complementary benefits of multi-view augmentation.

For example, on the Zhou2016 dataset, the best pairwise combination achieves an accuracy of 87.01\%, which is slightly lower than the 87.20\% obtained by MVCNet using all three views simultaneously. Similar trends can be observed on other datasets. These results highlight the necessity of incorporating multiple views to fully exploit the rich structure of EEG trials and improve model generalization.

\subsection{Feature Visualization}
\begin{figure*}[htpb]\centering
\subfigure[]{\includegraphics[width=\linewidth,clip]{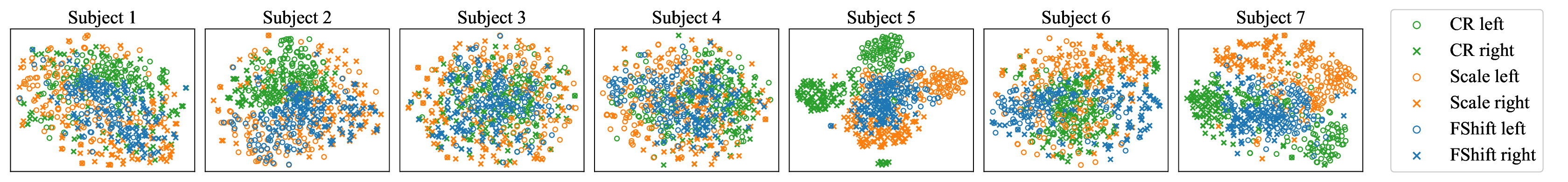}}
\subfigure[]{\includegraphics[width=\linewidth,clip]{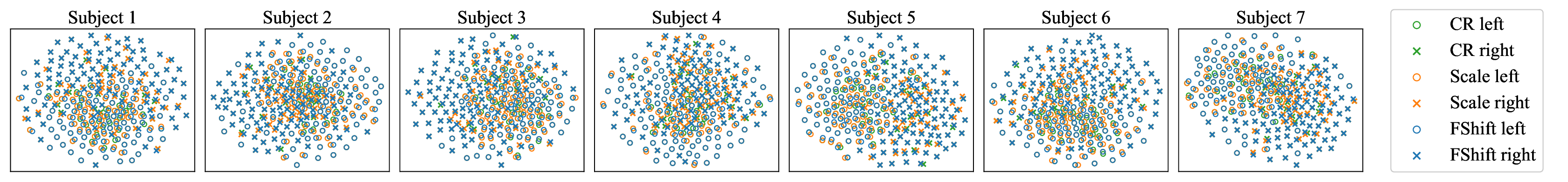}}
\caption{$t$-SNE visualizations of features extracted from seven subjects of the Blankertz2007 dataset: (a) EEGNet, and (b) MVCNet. Different augmentation types are encoded by colors, including CR (spatial domain), Scale (time domain), and FShift (frequency domain). Circles and crosses indicate two classes, respectively.} \label{fig:tsne}
\end{figure*}

$t$-SNE \cite{VanderMaaten2008} visualizations were conducted on features extracted from EEGNet and MVCNet under different augmentation strategies on the Blankertz2007 dataset, as shown in Figure~\ref{fig:tsne}. Compared with EEGNet, the feature clusters in MVCNet are more compact, with samples from different augmentation views better aligned within each class. This demonstrates the effectiveness of our cross-view contrastive module in bridging the gaps among time, spatial, and frequency domain transformations.

\subsection{Ablation Study and Parameter Sensitivity Analysis}
Ablation studies were conducted to evaluate the individual contributions of the two contrastive modules $\mathcal{L}_\textmd{CVC}$ and $\mathcal{L}_\textmd{CMC}$. As shown in Table \ref{tab:ablation}, performance improvements were observed when either module was included, while the combination of both consistently yielded the highest accuracy across all five datasets under the LOSO setting. These results suggest that the two losses are complementary and jointly beneficial for enhancing the performance.

\begin{table}[htbp] \centering \setlength{\tabcolsep}{1.5mm}
\small
\caption{Ablation study on $\mathcal{L}_\textmd{CVC}$ and $\mathcal{L}_\textmd{CMC}$ for all datasets.}
\begin{tabular}{cc|cccccc}
\toprule
$\mathcal{L}_\textmd{CVC}$ & $\mathcal{L}_\textmd{CMC}$ & D1 & D2 & D3 & D4 & D5 & Avg.\\
\midrule
\ding{55} & \ding{55} & 73.84 & 83.62 & 77.53 & 74.79 & 74.43 & 76.84 \\
\checkmark & \ding{55} & \underline{75.55} & \underline{86.36} & 78.59 & \underline{75.86} & \underline{75.08} & \underline{78.29} \\
\ding{55} & \checkmark & 74.95 & 85.20 & \underline{79.29} & 75.29 & 74.80 & 77.90 \\
\checkmark & \checkmark & \textbf{76.02} & \textbf{86.96} & \textbf{79.56} & \textbf{75.24} & \textbf{75.85} & \textbf{78.72} \\
\bottomrule
 \end{tabular}
\label{tab:ablation}
\end{table}

\begin{figure}[h]\centering
\includegraphics[width=\linewidth]{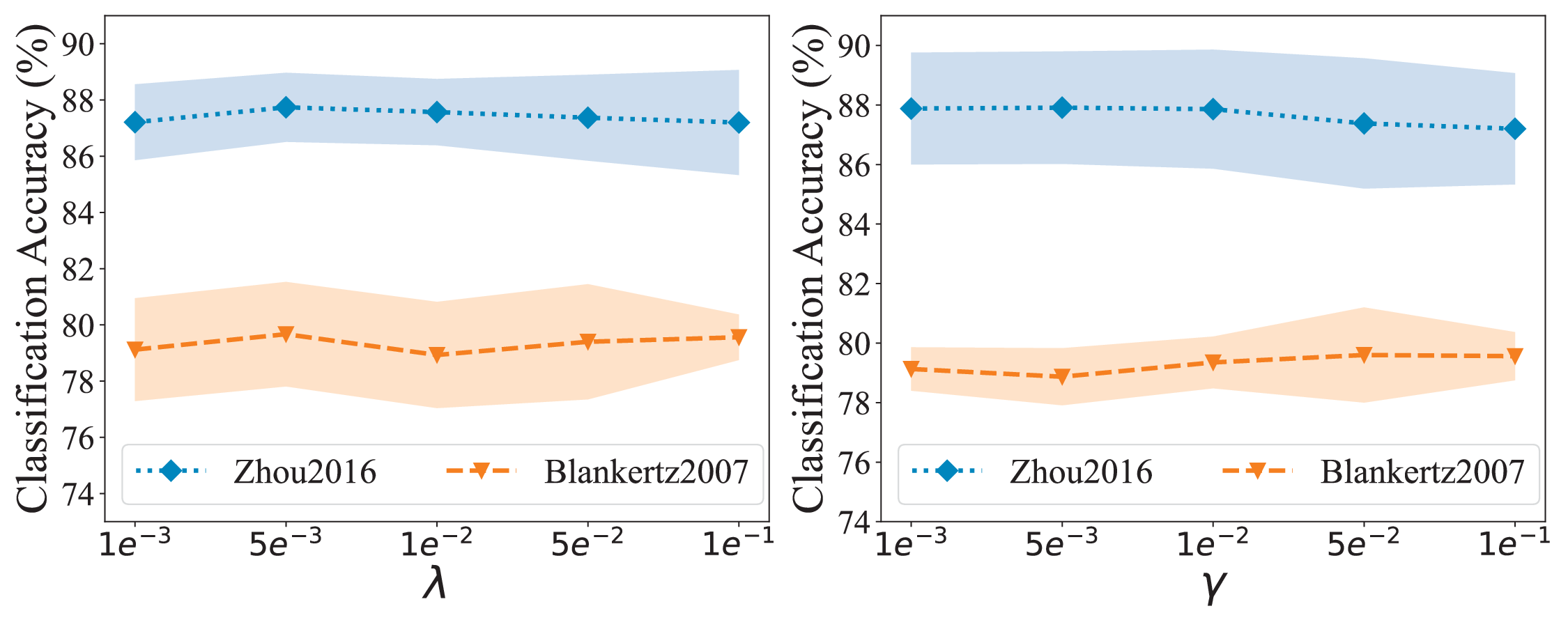}
\caption{Parameter sensitivity analysis of $\lambda$ and $\gamma$. When one parameter is varied, the other is fixed at 0.1. Each point indicates the average accuracy, and the shaded area represents the standard deviation.}
\label{fig:para_ana}
\end{figure}

To further assess the robustness of the proposed approach, sensitivity analyses were performed on the weighting parameters $\lambda$ and $\gamma$. In Figure~\ref{fig:para_ana}, one parameter was varied while the other was fixed at 0.1. Observe that the performance remained stable across a wide range of values.

\subsection{Effect of Transformer Heads and Layers}
\begin{figure}[h]\centering
\includegraphics[width=\linewidth]{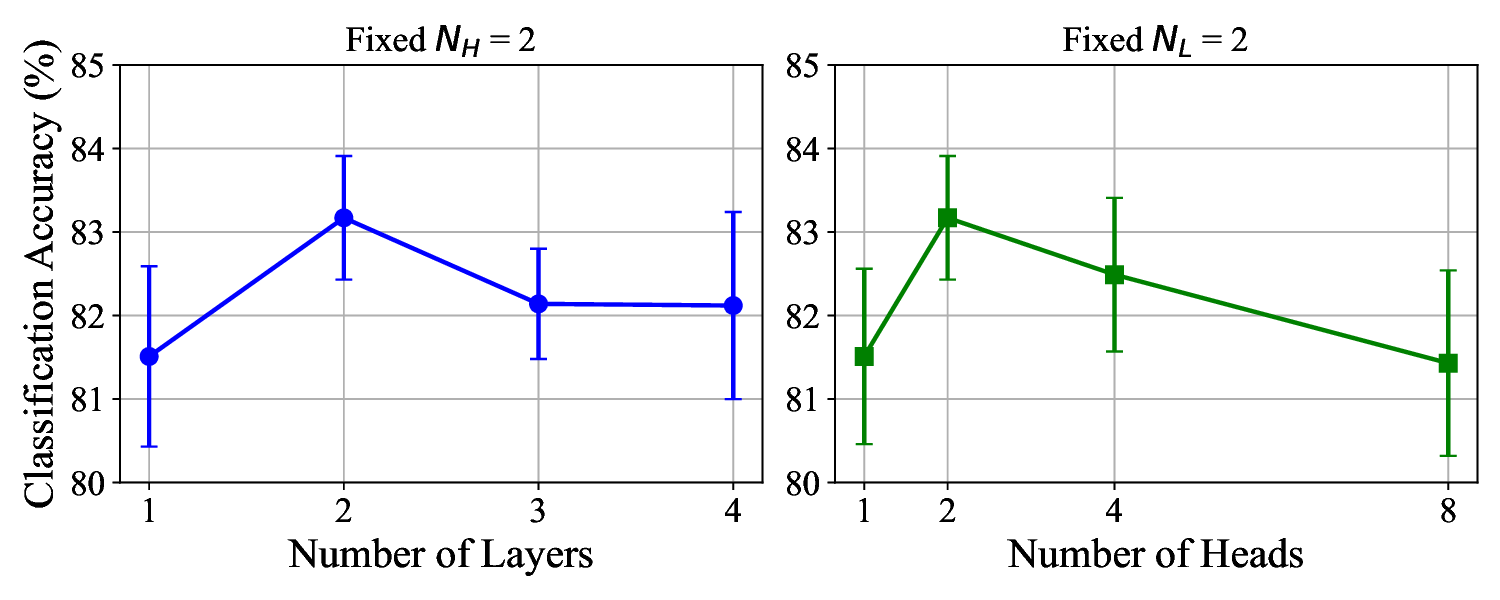}
\caption{Ablation study on the number of Transformer layers $N_L$ and heads $N_H$.}
\label{fig:ablation_nums}
\end{figure}

To evaluate the effect of the number of Transformer heads and layers, we conducted an additional ablation study by varying each parameter while keeping the other fixed. Specifically, we varied the number of Transformer layers $N_L \in \{1, 2, 3, 4\}$ with a fixed number of heads $N_H = 2$, and varied $N_H \in \{1, 2, 4, 8\}$ with $N_L = 2$ on BNCI2014001 dataset. Results are summarized in Figure~\ref{fig:ablation_nums}. Observed that:
\begin{itemize}
\item For the number of layers, increasing $N_L$ from 1 to 2 improves accuracy, but further increases to 3 or 4 do not yield additional gains and may even lead to slight degradation. This suggests that deeper attention structures may introduce overfitting or optimization challenges in small-scale EEG datasets.
\item For the number of heads, accuracy improves as $N_H$ increases from 1 to 2 or 4, indicating that multi-head attention enhances the model's ability to capture diverse temporal patterns. However, setting $N_H = 8$ leads to a slight performance drop, possibly due to redundant subspace partitioning or insufficient training data to support larger attention capacity.
\item Overall, the better performance is achieved with $N_L = 2$ and $N_H = 2$, which we adopt as the default configuration for MVCNet throughout the main experiments.
\end{itemize}

\section{Conclusion}\label{sect:conclusions}
This paper proposed MVCNet, a dual-branch network for MI decoding. By integrating CNN and Transformer modules in parallel, MVCNet effectively captures both local spatial-temporal features and global temporal dependencies. Cross-view and cross-model contrastive modules were introduced to enforce feature consistency across multiple views and network branches. Experimental results on five public MI datasets under three evaluation scenarios demonstrated that MVCNet consistently outperformed existing nine state-of-the-art models. Future research will explore more informative and effective model architectures, such as Mamba-based hybrid models, to further improve MI decoding performance.

\bibliographystyle{IEEEtran} \bibliography{scl_wang}

\end{document}